\documentclass[twocolumn,superscriptaddress,showpacs,amsmath,amssymb,aps]{revtex4-1}
\usepackage{graphicx}
\usepackage{amsmath}
\usepackage{multirow}
\usepackage{color}
\usepackage{bm}
\usepackage{braket}

\newcommand{\diff}{{\mathrm d}}
\newcommand{\rmc}{{\mathrm c}}

\newcommand{\hu}{h^\mathrm{u}}
\newcommand{\hs}{h^\mathrm{s}}

\usepackage[bookmarks=true,bookmarksnumbered=true,bookmarkstype=toc,pdfborder={0 0 0},citecolor=blue,linkcolor=blue,colorlinks=true]{hyperref}

\begin{document}

\title{Stochastic approximation of dynamical exponent at quantum critical point}

\author{Shinya Yasuda}
\affiliation{Department of Applied Physics, University of Tokyo, Tokyo 113-8656, Japan}
\author{Hidemaro Suwa}
\affiliation{Department of Physics, University of Tokyo, Tokyo 113-0033, Japan}
\author{Synge Todo}
\affiliation{Department of Physics, University of Tokyo, Tokyo 113-0033, Japan}
\affiliation{Institute for Solid State Physics, University of Tokyo, Kashiwa, 277-8581, Japan}

\date{\today}

\begin{abstract} 
  We have developed a unified finite-size scaling method for quantum phase transitions that requires no prior knowledge of the dynamical exponent $z$.  During a quantum Monte Carlo simulation, the temperature is automatically tuned by the Robbins-Monro stochastic approximation method, being proportional to the lowest gap of the finite-size system.
  The dynamical exponent is estimated in a straightforward way from the system-size dependence of the temperature. As a demonstration of our novel method, the two-dimensional $S=1/2$ quantum $XY$ model in uniform and staggered magnetic fields is investigated in the combination of the world-line quantum Monte Carlo worm algorithm. In the absence of the uniform magnetic field, we obtain the fully consistent result with the Lorentz invariance at the quantum critical point, $z=1$, i.e., the three-dimensional classical $XY$ universality class. Under a finite uniform magnetic field, on the other hand, the dynamical exponent becomes two, and the mean-field universality with effective dimension $(2+2)$ governs the quantum phase transition.
\end{abstract}


\maketitle

\section{Introduction}\label{introduction}
Recent enhancement of the computational power has enabled us to
simulate larger-scale systems with higher precision than ever
before. In particular, with the help of the recent development of simulation
algorithms for strongly correlated quantum systems, a number of simulations have been performed to elucidate
the novel nature of quantum phase transitions, in which many-body
physics plays an essential
role~\cite{Sachdev1999,AvellaM2013}. Quantum phase transitions occur
at absolute zero temperature, triggered by quantum
fluctuations.  Through the quantum-classical mapping, a quantum phase
transition in $d$ dimensions, if it is of second order, can be generally described by the
critical theory as the temperature-driven phase transition in a
$(d+z)$-dimensional classical system with the same symmetries, where $z$
is the so-called dynamical exponent~\cite{Suzuki1976,Vojta2003}.

A world-line quantum Monte Carlo (WLQMC) method is one of the most powerful tools for investigating quantum critical phenomena without any bias or approximation~\cite{Suzuki1994,LandauB2005}. A quantum system in $d$ dimensions is mapped to a classical system in $(d+1)$ dimensions in the WLQMC method.  The system length along the additional direction, the imaginary-time direction, is given by the inverse temperature, $\beta$.

When one performs a WLQMC simulation to investigate quantum criticality, the choice of $\beta$ for each system size is essential. The reason is that the quantum critical system can be extremely anisotropic even if the interactions are isotropic in real space. While the correlation length in the real-space direction diverges as $\xi\sim (g-g_\rmc)^{-\nu}$, that in the imaginary-time direction does as $\xi_\tau\sim (g-g_\rmc)^{-z\nu}$, where $g$ is the coupling constant that controls quantum fluctuations, $g_\rmc$ the quantum critical point, and $\nu$ the critical exponent.  If the dynamical exponent $z$ is one, the space-time isotropy is kept aside from a scale factor, or the velocity of low-energy excitation. In the meanwhile, there are phase transitions with a dynamical exponent larger than one. The Bose-Hubbard model with randomness that exhibits quantum criticality with $z>1$ has been extensively investigated analytically~\cite{FisherWGF1989, SinghR1992, Herbut1997, WeichmanM2007}, numerically~\cite{SorensenWGY1992, ZhangKCG1995, ProkofievS2004, PriyadarsheeCLB2006, MeierW2012, ZunigaLLL2015} as well as experimentally~\cite{HuvonenZMYRNLGZ2012,YuMWMPZR2012}.

Let us review the renormalization group and the scaling theory near a quantum critical point.  Consider the scale transformation with a certain length scale, $b$.
A physical quantity, denoted as $F$, is generally transformed as 
\begin{equation}
  \label{fss}
  \begin{split}
F(g-g_\rmc, L^{-1}, \beta^{-1})&=b^{y_F}F(b^{1/\nu}(g-g_\rmc), bL^{-1}, b^{z}\beta^{-1}) \\
&= L^{y_F} \tilde{F}(L^{1/\nu}(g-g_\rmc),L^z\beta^{-1}),
  \end{split}
\end{equation}
where $y_F$ is the scaling dimension of the quantity under
consideration. In the second line of Eq.\,\eqref{fss}, we chose $b=L$
and introduced $\tilde{F}(x,y) \equiv F(x, 1, y)$.  This equation
has several unknown constants, $g_\rmc$, $y_F$, $\nu$, $z$, and
the scaling function $\tilde{F}(x,y)$ itself.

In order to
determine the constants in the finite-size scaling
ansatz~\eqref{fss}, one had to repeat simulations densely in the
three-dimensional parameter space~$(L,\beta,g)$, and perform a
multi-parameter finite-size scaling analysis as in
Refs.\,\citenum{RiegerY1994} and \citenum{PichYRK1998}. It typically
requires considerable computational resources to scan the
multi-dimensional parameter space. Instead of the exhaustive scanning,
simulations with some assumed $z$ were performed in most previous studies. The consistency was checked after the calculation as in Ref.\,\citenum{Pollet2013}. This approach, however, would be awfully
inefficient in the case without knowledge of the value of $z$ in advance. Another
approach for $z$ estimation was to focus on the temperature dependence of the
correlation length, $\xi \sim \beta^{1/z}$ at $g=g_\rmc$ and $L=\infty$~\cite{PriyadarsheeCLB2006}. After the correlation length in the thermodynamic limit was extrapolated at each temperature, the low-temperature asymptotic behavior was analyzed. This two-step procedure requires additional
computational cost, and possibly introduces some uncontrollable
systematic error from the extrapolations, even if the location of quantum critical point, $g_\rmc$, is known.  In the meantime, the winding numbers of the world-lines in space and time directions were exploited in Refs.\,\citenum{Jiang2011}
and \citenum{Jiang2012}. A parameter, $L$ or $\beta$, was
interpolated so that the winding number squared in each direction
averagely took the same value.
However, such an interpolation is again non-trivial in a multi-parameter space and multi conditions.

\begin{figure}[tbp]
\begin{center}
\vspace{1em}
\includegraphics[width=85mm]{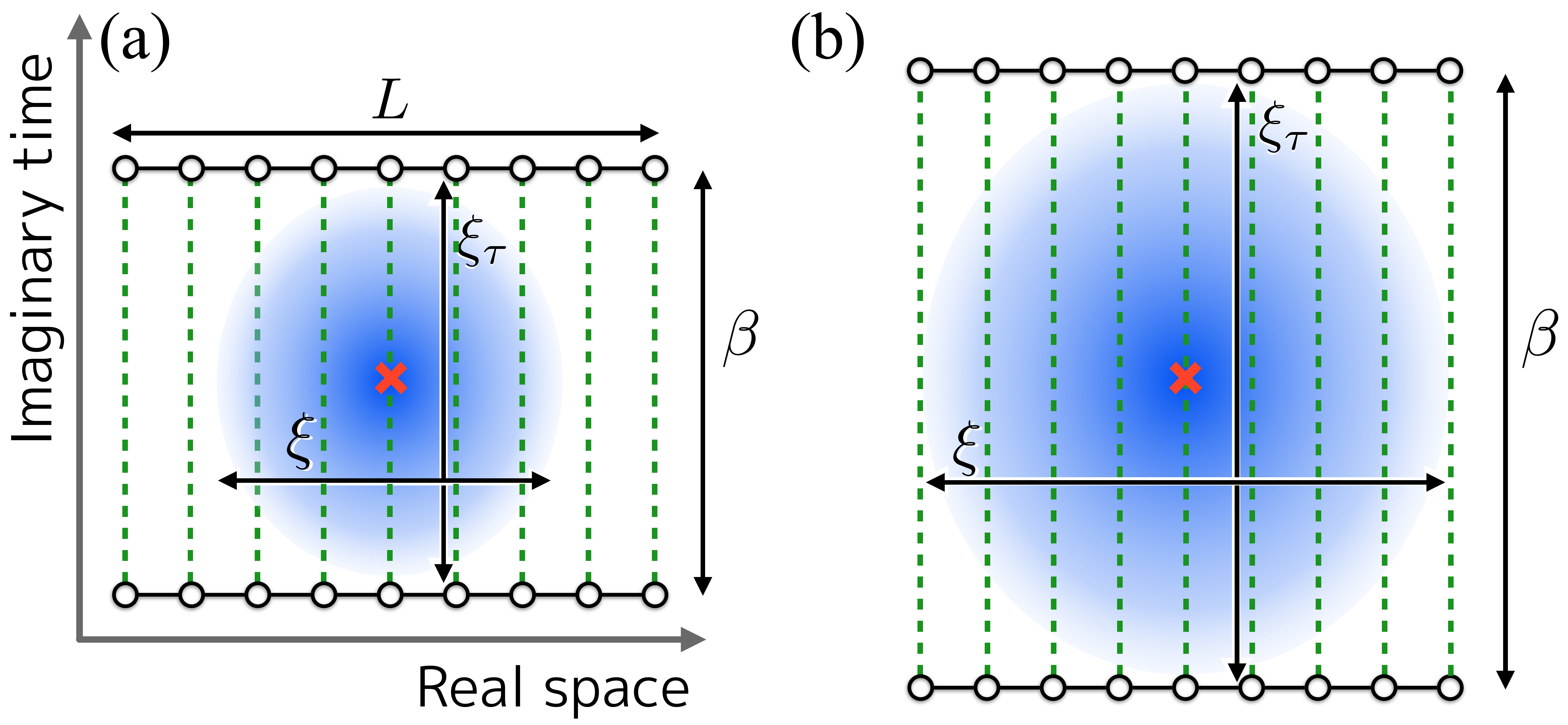}
\vspace{-1em}
\caption
  {(Color online) Schematic pictures of virtually (a) anisotropic and (b) isotropic systems, where $L$ and
  $\beta$ are the system linear length in the real space and the imaginary time
  directions, respectively.
  The blue oval in each picture depicts space-time region correlated with the center (red cross).
  After the aspect-ratio optimization, the relative correlation lengths, $\xi/L$ and $\xi_\tau/\beta$, become almost the same.}
\label{control.pdf}
\end{center}
\end{figure}

One of the most effective strategies to overcome the difficulty of a
multi-parameter scaling is to introduce an auto-tuning technique.
A number of auto-tuning techniques have already been used in numerical simulations in the field of statistical physics. For example, the invaded cluster algorithm \cite{MachtaCLSC1995}
or the probability-changing cluster algorithm \cite{TomitaO2001}
can automatically locate the critical point. The Wang-Landau algorithm \cite{WangL2001a}
enables us to directly estimate the density of states of a system.

In the present paper, we employ the stochastic approximation method.
Recently in Ref.\,\citenum{YasudaT2013}, a
method to automatically optimize the aspect ratio of a quantum system
was proposed for analyzing quantum criticality under strong spatial anisotropy. The relative correlation length,
$R_\alpha\equiv\xi_\alpha/L_\alpha$, where $\xi_\alpha$ and $L_\alpha$ are the
correlation length and the system size in $\alpha$ direction, respectively ($\alpha=x$, $y$, or $\tau$
for two-dimensional systems and $L_\tau=\beta$), was adjusted for making the system virtually isotropic as
$R_x:R_y:R_\tau \approx 1:1:1$. Figure\,\ref{control.pdf}
schematically illustrates virtually anisotropic and isotropic
one-dimensional quantum systems.  In Ref.\,\citenum{YasudaT2013}, the
stochastic approximation scheme was applied to the staggered dimer
antiferromagnetic Heisenberg model, and 
the universality class of the quantum critical point was successfully identified in spite of the
existence of strong finite-size corrections that easily lead a naive finite-size scaling analysis to an incorrect conclusion~\cite{WenzelJ2009,WenzelBJ2008}. Note that the tuning method using the correlation length~\cite{YasudaT2013} is applicable to general systems, while the
method based on the winding number\,\cite{Jiang2011,Jiang2012} works only for
systems with U(1) symmetry.

The aim of the present paper is to propose a unified finite-size scaling method based on the stochastic approximation technique for quantum criticality with general $z$. The relevant critical exponents including the dynamical exponent will be obtained simultaneously without any prior knowledge or assumption of the values.
We will demonstrate our approach for the two-dimensional $S=1/2$ quantum $XY$ model in
uniform and staggered magnetic fields along $z$ direction (in spin
space). We will clarify that the dynamical exponent becomes two under a finite uniform magnetic field, while it does one in the absence of a uniform field.

This paper is organized as follows: Sec.\,\ref{method}
introduces the scaling ansatz of the correlation lengths for
space-time anisotropic systems, the Robbins-Monro stochastic
approximation method, and its convergence property.  It is also discussed how
the stochastic approximation method is applied to the present
finite-size scaling analysis. In Sec.\,\ref{model}, the model
considered in the present paper and the WLQMC method are
introduced. The numerical results are shown in
Sec.\,\ref{results}. Finally our study is concluded in Sec.\,\ref{summary}.  The
technical details are reported in Appendices~\ref{RM_convergence} and \ref{worm}.

\section{Stochastic approximation method}\label{method}

\subsection{Conditions for realizing space-time isotropy} 

As noted in Sec.\,\ref{introduction}, for the system with $z>1$, one should pay
attention to the space-time aspect ratio when considering the finite-size
scaling analysis.  In this section, we explain conditions to realize a virtually isotropic system during a simulation.  For simplicity, we assume that the model
considered hereafter has no anisotropy in real space. Generalization to systems with spatial anisotropy is straightforward.

Let us start by choosing $F=\xi$ in Eq.\,\eqref{fss}. In this case, $y_F$ is one, i.e.,
\begin{equation}
\xi(g-g_\rmc,L^{-1},\beta^{-1})=L\,\tilde{\xi}((g-g_\rmc)L^{1/\nu},L^z/\beta).\label{fss_xi}
\end{equation}
Another choice, $F=\xi_\tau$, yields
\begin{equation}
\xi_\tau(g-g_\rmc,L^{-1},\beta^{-1})=L^{y_\tau}\tilde{\xi_\tau}((g-g_\rmc)L^{1/\nu},L^z/\beta),\label{fss_xi_tau}
\end{equation}
where we set $y_F = y_\tau$. At the quantum critical point, $g=g_\rmc$, the correlation length in the imaginary-time direction exhibits the power law, $\xi_\tau(L^{-1})\propto L^{y_\tau}$, in the limit of
$\beta\to\infty$.  By the definition
of the dynamical exponent, one finds $y_\tau=z$. 
Then, dividing both  sides of Eq.\,\eqref{fss_xi_tau} by $\beta$ yields
\begin{equation}
\xi_\tau(g-g_\rmc,L^{-1},\beta^{-1})/\beta=\tilde{\xi}_\tau^\prime((g-g_\rmc)L^{1/\nu},L^z/\beta),\label{fss_xi_tau2}
\end{equation}
where $\tilde{\xi}_\tau^\prime (x,y) \equiv y \, \tilde{\xi}_\tau (x,y)$.

Here, let us introduce two conditions,
\begin{equation}
  \xi/L=R,
  \label{eqn:condition1}
\end{equation}
where $R$ is an arbitrarily chosen constant, and
\begin{equation}
  \xi_\tau/\beta=R_\tau,
  \label{eqn:condition2}
\end{equation}
where $R_\tau$ is another constant. 
Assume that conditions~\eqref{eqn:condition1} and \eqref{eqn:condition2} are both satisfied by tuning $g$ and $\beta$
in simulating systems with different system sizes.  If this is the case,
$\tilde{\xi}$ and $\tilde{\xi}_\tau^\prime$ are kept constant even though they are different functions.
Meanwhile, the set of arguments of $\tilde{\xi}$ in Eq.\,\eqref{fss_xi} and that of $\tilde{\xi}_\tau^\prime$ in Eq.\,\eqref{fss_xi_tau2} are the same as each other. That is, the different functions $\tilde{\xi}$ and $\tilde{\xi}_\tau^\prime$ sharing the same arguments are kept constant at different system sizes. This means that each of the arguments
should be constant if the functions have some
reasonable monotonicity. The monotonicity of the scaling functions is expected to hold near a generic critical point and supported by our numerical calculation shown below. Then Eqs.~\eqref{eqn:condition1} and \eqref{eqn:condition2} provide solutions, $g_\rmc(L)$ and $\beta(L)$, for each system size:
\begin{align}
  &g_\rmc(L) - g_\rmc \propto L^{-1/\nu} \label{critical_point}
\end{align}
and
\begin{align}
&\beta(L) \propto L^{z}\label{dynamical_exponent}.
\end{align}
Thus, the coupling constant, $g$, automatically converges to the critical point as $L$ increases. Moreover, the dynamical exponent can be simultaneously estimated from the asymptotic $L$ dependence of the inverse temperature, $\beta$.

\subsection{Robbins-Monro stochastic approximation method}
In this section, we introduce an iteration procedure to fulfill the conditions proposed in the previous
section.  Our task is to solve the system of nonlinear equations, $\xi/L=R$ and
$\xi_\tau/\beta=R_\tau$, with respect to $g$ and $\beta$ for given $L$, $R$, and
$R_\tau$.  The solution cannot be obtained by standard iterative methods for nonlinear
equations, such as the Newton-Raphson method. It is because $\xi$ and $\xi_\tau$ have statistical errors coming
from the Monte Carlo sampling that make the conventional methods unstable.  We thus employ
the stochastic approximation method explained below. 

Let us see a concrete example of the stochastic approximation.  For simplicity, assume that $g$
is already set to $g_\rmc$. We estimate the
optimal $\beta$ that satisfies the relation $\xi_\tau/\beta=R_\tau$. 
The solution of this equation is denoted as $\beta_\rmc$.
First, one runs a short Monte Carlo simulation with a trial parameter $\beta^{(1)}$ and measures the correlation
length, then calculates
$A(\beta^{(1)}) \equiv R_\tau-\xi_\tau/\beta^{(1)}$.  Next, one updates the
parameter, $\beta$, by using the Robbins-Monro type update procedure~\cite{RobbinsM1951,
Bishop2006}
\begin{equation}
\beta^{(n+1)}=\beta^{(n)}-\frac{p}{n}A(\beta^{(n)})\label{update}
\end{equation}
with $n=1$ and repeats the above until $\beta^{(n)}$ converges to a
certain value with increasing $n=2,3,4,\cdots$.  Here, $p$ is a (constant)
parameter that determines the gain of the feedback.  Regardless
of the choice of the gain, it is proved that $\beta^{(n)}$ converges
to $\beta_\rmc$ in $n\to\infty$ with probability one~\cite{RobbinsM1951, AlbertG1970}.

As explained in Appendix~\ref{RM_convergence}, the mean of the
probability distribution of $\beta^{(n)}$ at the $n$-th step (denoted
as $\mu_n$) converges as $\mu_n - \beta_\rmc \sim 1/n^{ap}$, where $a$
is the derivative of $A(\beta)$ at $\beta=\beta_\rmc$, and the sign of
$p$ is chosen as the same with $a$. For $ap \leq 1/2$, the variance of
$\beta^{(n)}$ at the $n$-th step (denoted as $\sigma_n^2$) is
evaluated as $\sigma_n^2\sim 1/n^{2ap}$. For $ap > 1/2$, on the other
hand, $\sigma_n^2\sim s^2/n$, where $s^2\equiv\sigma^2p^2/(2ap-1)$ is
the asymptotic variance and $\sigma$ is the statistical error
resulting from a Monte Carlo estimation of $A(\beta)$. Here we should set $p
\approx 1/a$ to minimize the variance (see the detailed discussion in
Appendix~\ref{RM_convergence}).
By this choice of $p$, it is also guaranteed that the systematic error of $\beta^{(n)}$ decreases faster than the statistical
(standard) error.
In actual simulations, one needs to
perform some ($\sim$10 at least) independent stochastic
approximation processes to estimate error bars of $\beta$ and physical quantities. The number of steps of
each approximation process has to be large enough ($\gtrsim10^2$
typically) for the systematic error to become negligible in comparison to the statistical error.

The present stochastic approximation method can be extended to multi-dimensional problems in a straightforward way.
Below, we will apply the method to the quantum phase transition of two-dimensional $S=1/2$ $XY$ model in uniform and staggered magnetic fields in order to demonstrate the efficiency of the present approach and clarify the quantum phase transitions.

\section{Model and quantum Monte Carlo method}\label{model}
\subsection{$S=1/2$ quantum $XY$ model in uniform and staggered magnetic fields}
The Hamiltonian of the two-dimensional $S=1/2$ quantum $XY$ model in uniform and staggered magnetic fields is defined as follows:
\begin{equation}
\mathcal{H}=-\frac{1}{2}\sum_{\langle j,k\rangle}(S_j^+S_k^-+S_k^+S_j^-)-\sum_j \left[ \hu + \hs(-1)^{\sigma(j)}\right]S_j^z,\label{hamiltonian}
\end{equation}
where $S_j^+$ ($S_j^-$) is the $S^z$-component raising (lowering) operator at site $j$, $\langle j,k\rangle$ denotes a pair of nearest-neighboring spins, and $\hu$ ($\hs$) is the amplitude of the uniform (staggered) magnetic field.
Here we consider the square lattice of linear extent $L$ with the periodic boundary conditions, and
the lattice is bipartite with even $L$. If site $j$ belongs to one of the sublattices,
$\sigma(j)$ takes zero, otherwise $\sigma(j)=1$. 

This model can be mapped to the
hard-core boson model with the uniform and the staggered chemical
potentials~\cite{LiebSM1961}.
The phase diagram of the
model consists of several phases~\cite{HenIR2010,HenR2009}: (i) the disordered phase that corresponds to the insulating or pinning phase in the boson model, (ii) the $xy$-plane ferromagnetic phase with non-zero transverse magnetization, or the compressible superfluid
phase, and (iii) the fully-polarized phase along $\hu$, or the empty (fully-occupied) phase.  We will fix $\hu$ to some value and change $\hs$ across the phase boundary.  When $\hu$ is smaller than the saturation field, $\hu_\rmc = 2$, a phase
transition from the ferromagnetic phase to the disordered phase occurs as
$\hs$ increases. If $\hu$ is larger than $\hu_\rmc$, an additional
phase transition from the fully-polarized phase to the ferromagnetic phase
occurs. When $\hu=0$, the particle-hole
symmetry holds and the phase transition is known to belong to the
three-dimensional $XY$ (3D-$XY$) universality class, i.e., $z=1$. 
Phase transitions different from the 3D-$XY$ universality with $z>1$, on
the other hand, are expected for $\hu\neq 0$~\cite{HenIR2010,
  FisherWGF1989}.

\subsection{World-line quantum Monte Carlo worm algorithm}
In order to simulate the system described by Hamiltonian~\eqref{hamiltonian}, we used the worm (directed-loop) algorithm~\cite{ProkovievST1998, SyljuasenS2002, KawashimaH2004} with the continuous-time path-integral representation. In the continuous-time representation, we introduce imaginary time $\tau$ as
\begin{align}
Z=\mathrm{Tr}\,\mathrm{e}^{-\beta\mathcal{H}}
=\mathrm{Tr}\left[\exp\left(-\int_0^\beta \!\! \mathrm{d}\tau \, \mathcal{H}\right)\right],\label{partition1}
\end{align}
where $Z$ is the partition function. The continuous-time formulation
was adopted because of the convenience for calculating the Fourier component of the imaginary-time correlation function, which we
will exploit to calculate $\xi_\tau$.  Expanding the exponential in the r.h.s.~of
Eq.\,\eqref{partition1}, we insert the identity, $\sum_m
\ket{\phi_m}\bra{\phi_m} = 1$, between the operators, where $\{ \ket{\phi_m} \}$ is a complete basis set of the Hilbert space. We then obtain
\begin{align}
Z=&1+\sum_{n=1}^{\infty}\sum_{(\phi_1,\dots ,\phi_n)} \int_0^{\beta}\!\!\diff\tau_1\cdots\int_{\tau_{n-1}}^\beta\!\!\!\!\diff\tau_{n}\nonumber\\
&\qquad \qquad \times \prod_{\ell=1}^n\bra{\phi_\ell}(-\mathcal{H})\ket{\phi_{\ell+1}},\label{partition3}
\end{align}
where $\ket{\phi_{n+1}}=\ket{\phi_1}$. 
In our WLQMC simulation, a state in the basis set is the direct product of the eigenstate of the local $S^z$ operator (up or down).
The Hamiltonian~\eqref{hamiltonian} conserves total $S^z$ of the system and thus 
a space-time configuration forms continuous lines of up spins (or down spins), i.e., the world-lines.

One can consider the integrand in Eq.\,\eqref{partition3} as a weight (probability measure) of each world-line configuration.
In order to make a simulation efficient, the second (site) term in
Hamiltonian~\eqref{hamiltonian} is included in the bond term as
\begin{equation}
\frac{1}{4}\sum_{\langle j,k\rangle}\left[\hu(S_j^z+S_k^z)+\hs(-1)^{\sigma(j)}(S_j^z-S_k^z)\right],
\end{equation}
where the factor $1/4$ comes from the coordination number of the square lattice. 
The matrix elements of the combined bond term are expressed as
\begin{equation}
\begin{bmatrix}
\hu/4  & 0                       & 0    & 0 \\
0      & \hs(-1)^{\sigma(j)}/4    & 1/2 & 0 \\
0      & 1/2    & -\hs(-1)^{\sigma(j)}/4 & 0 \\
0      & 0 & 0                    & -\hu/4
\end{bmatrix}
\begin{matrix}
\ket{\uparrow\uparrow} \\
\ket{\uparrow\downarrow} \\
\ket{\downarrow\uparrow} \\
\ket{\downarrow\downarrow}
\end{matrix}
\label{bondweight} .
\end{equation}
A constant larger than or equal to $\max( |\hu|/4, |\hs|/4 )$ needs to be added to the diagonal elements for ensuring the non-negativity of the weights.

In the worm algorithm, extended world-line configurations are
introduced.  The configurations to sample in the Monte Carlo method consist of the original world-lines and the
world-lines with a pair of kinks, points of discontinuity.  Such a pair is called a
\textit{worm}, and each of discontinuity \textit{head} or \textit{tail}.  In the present spin model, the worm is represented by the pair of spin ladder operators, ($S_j^+$, $S_k^-$) or ($S_j^-$,$S_k^+$), each of which is defined at a space-time point.

\begin{figure}[tbp]
\centering
\vspace{1em}
\includegraphics[width=85mm]{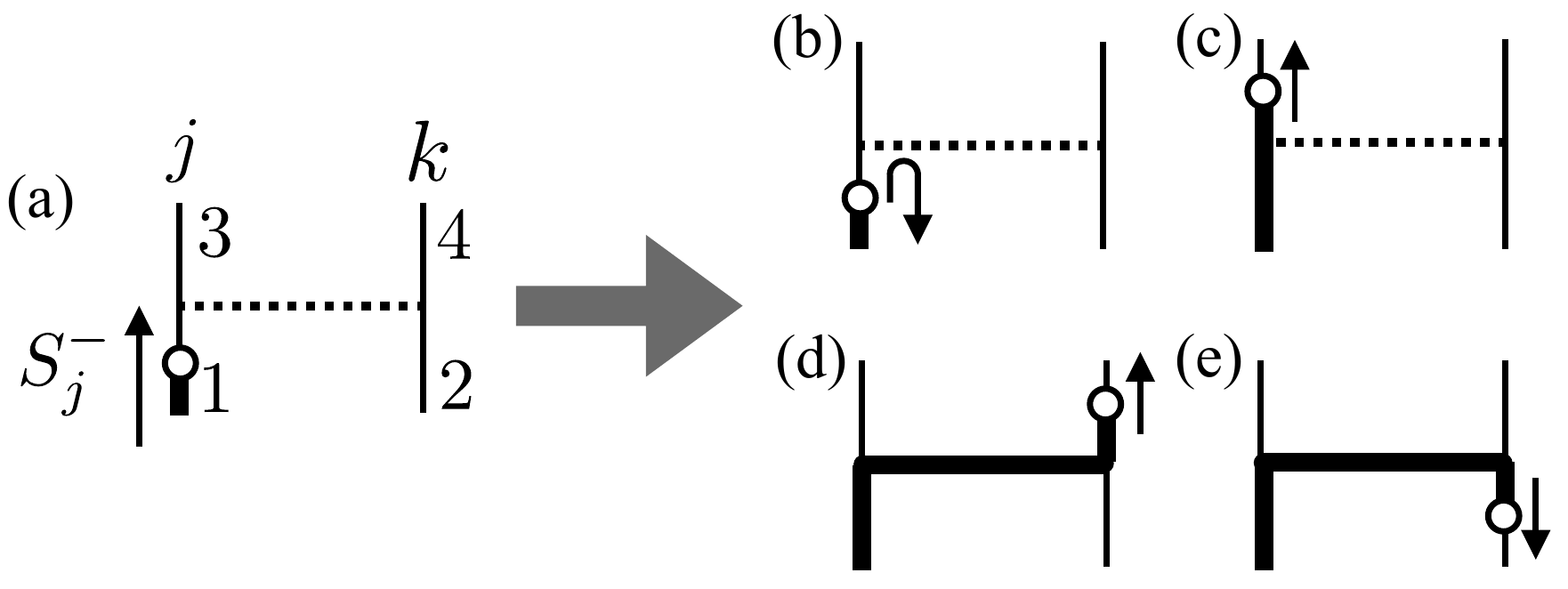}
\caption{Example of the worm scattering process. The bold line denotes a string of $S^z=1/2$ states (or the path of the worm head), while the thin line does $S^z=-1/2$ states. The dotted line expresses a bond operator. When the worm head, $S_j^-$, (a)~arrives at leg 1 of the bond operator, one of the following events will occur. The head (b)~bounces back to the way which it comes from, (c)~goes straight and gets out from leg 3, (d)~jumps to leg 4, or (e)~turns to leg 2. The transition probabilities in the scattering process are determined by the matrix elements\,\eqref{bondweight}. In the present model, event (e) never occurs due to the total $S^z$ conservation.} \label{update.pdf}
\end{figure}

The whole update process of the world-line configuration consists of the diagonal update and the worm update~\cite{KawashimaH2004}. In the former, the
diagonal bond operators in the Hamiltonian~\eqref{hamiltonian} are inserted into or
removed from a world-line configuration according to the diagonal elements. In the latter, first a worm,
i.e., a pair of the raising and lowering operators, is inserted at a
randomly chosen space-time point, and either operator is chosen as the head. The order of the ladder operators is uniquely determined in the case with $S=1/2$ since the local degree of freedom is binary (up or down). The worm head then moves along
the imaginary-time direction until it arrives at a bond operator. At
the operator, the worm head is scattered and its moving direction and/or sitting site may be changed stochastically according to the matrix elements\,\eqref{bondweight}. In Fig.\,\ref{update.pdf}, an
example of the worm-scattering process is illustrated. We choose transition probabilities so as
to minimize the bounce probability (see Fig.\,\ref{update.pdf}), by breaking both the detailed
balance of each worm-scattering process and even that of the whole
Monte Carlo dynamics~\cite{SuwaT2010}. This scattering process is repeated until the
worm head reaches back its own tail. Then the head and tail destroy each other. The worm is inserted at several times in each Monte Carlo step.

We will investigate the phase transition between the $xy$-plane ferromagnetic phase and the disordered phase. The order parameter is the 
transverse (off-diagonal) magnetization in $x$ or $y$ direction.
Although it is non-trivial to measure off-diagonal
correlation in a WLQMC simulation, one can
efficiently calculate the structure factor
\begin{equation}
S_0 =\frac{1}{L^2}\Big\langle\sum_{j,k} (S_j^x S_k^x + S_j^y S_k^y) \Big\rangle\label{def_sfac} ,
\end{equation}
the transverse susceptibility
\begin{equation}
  \chi = \frac{1}{L^2\beta} \Big\langle\sum_{j,k}\int_0^\beta\!\!\diff\tau_1 \int_0^\beta\!\! \diff\tau_2 \, S_j^+(\tau_1)S_k^-(\tau_2)\Big\rangle ,
  \label{def_sus}
\end{equation}
and the Fourier component of the (spatial and temporal) correlation
functions, exploiting the virtue of the worm-update
process~\cite{ProkovievST1998}. Here the symbols are defined as follows: $\langle O \rangle = \mathrm{Tr} [ O \mathrm{e}^{-\beta H} ]/Z$ and $O(\tau)=\mathrm{e}^{-\tau H} O \mathrm{e}^{\tau H}$. The correlation lengths in $x$, $y$,
and $\tau$ directions are then calculated from the Fourier components
by the second-moment method~\cite{CooperFP1982,TodoK2001}.  The detail
of the measurement of these quantities is explained in
Appendix~\ref{worm}.

At the critical point, the structure factor and
the susceptibility exhibit the following power-law behavior:
\begin{equation}
S_0(L)\propto L^{\theta},\label{s_fac}
\end{equation}
and 
\begin{equation}
\chi(L) \propto L^{\gamma/\nu},\label{susceptibility}
\end{equation} 
respectively.
Here we introduce $\theta \equiv 2-2{\beta}/\nu$, where $\beta$ is not
the inverse temperature but the critical exponent of the order
parameter.  The exponent of the susceptibility is denoted by $\gamma$. Note that in Eqs.\,\eqref{dynamical_exponent}, \eqref{s_fac}, and \eqref{susceptibility}, one can use the quantities evaluated at $g_\rmc(L)$, the solution of Eqs.~\eqref{eqn:condition1} and \eqref{eqn:condition2} for each system size $L$, instead of the true critical point $g_\rmc$, as both give the same exponent.

\section{Numerical results}\label{results}
\subsection{For $\hu=0$}
First we discuss the case without the uniform magnetic field.
We performed WLQMC simulations for system sizes $L=8,\cdots,64$, and obtained
the optimal inverse temperature $\beta(L)$ and staggered magnetic field $\hs_\rmc(L)$ for each $L$ by solving Eqs.\,\eqref{eqn:condition1} and \eqref{eqn:condition2} using the stochastic approximation. The optimal inverse temperature $\beta(L)$ ensures that the system is virtually isotropic, and the optimal staggered magnetic field $\hs_\rmc(L)$ gives an estimate of the critical point. The target relative correlation lengths, $R$ and $R_\tau$, were chosen as $R=R_\tau=0.5$, $R_{XY}$, or 0.7, where $R_{XY} = 0.5925$ is the approximate value of $\lim_{L \rightarrow \infty} \xi(L)/L$ at the critical point of the three-dimensional classical $XY$ model~\cite{CampostriniHPRV2001}.  Note that a particular choice of $R$ and $R_\tau$ does not introduce any bias to the final estimates; it affects only the speed of convergence to the thermodynamic limit as seen below.

\begin{figure}[tbp]
\centering
\vspace{1em}
\includegraphics[width=80mm]{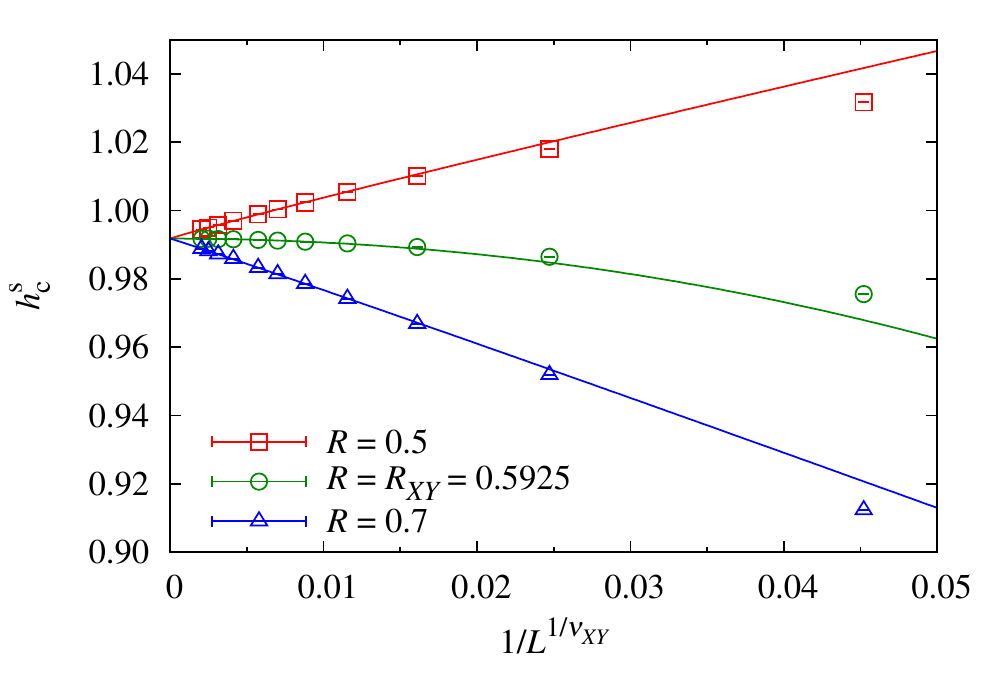}
\vspace{-1em}
\caption
  {(Color online) System-size dependence of the critical staggered field, $\hs_\rmc(L)$, for $\hu=0$. The extrapolated value is $\hs_\rmc=0.99179(3)$, and the lines are the bootstrapped fitting curves (see the body).}
\label{h0_A_2.pdf}
\end{figure}

\begin{figure}[tbp]
\centering
\vspace{1em}
\includegraphics[width=80mm]{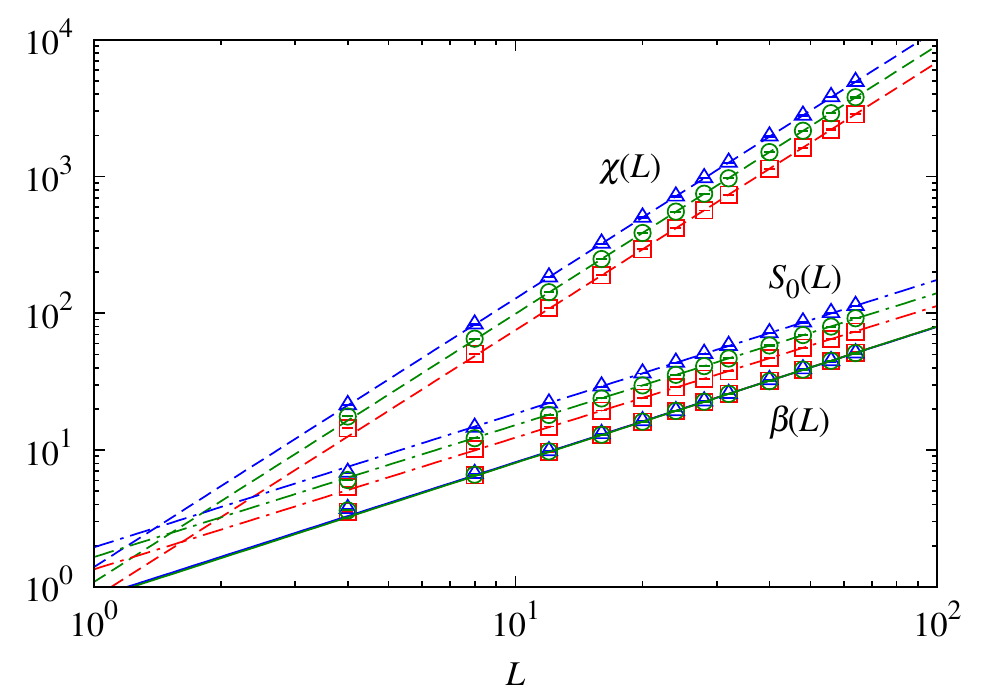}
\vspace{-1em}
\caption
  {(Color online) System-size dependence of the transverse susceptibility\,\eqref{def_sus}, the structure factor\,\eqref{def_sfac}, and the inverse temperature at $\hs_\rmc(L)$ for $\hu=0$. The red squares, green circles, and blue triangles are the data plots for $R=0.5$, $R_{XY}$, and $0.7$, respectively.}
\label{h0_all.pdf}
\end{figure}

The critical strength of the staggered magnetic field, $\hs_\rmc$, can be estimated based on the asymptotic form~\eqref{critical_point}, i.e.,
\begin{equation}
\hs_\rmc(L)=\hs_\rmc+c(R) L^{-1/\nu},\label{extrapolate_A}
\end{equation}
where $c(R)$ is a certain constant which depends only on $R$
($=R_\tau$, here).  The system-size dependence of $\hs_\rmc(L)$ is
shown in Fig.\,\ref{h0_A_2.pdf}.  We extrapolated the critical point
from the finite-size data by using the ansatz
Eq.\,\eqref{extrapolate_A} and assuming the same $\hs_\rmc$ for three
different values of $R$.  Seven parameters, $\hs_\rmc$ and ($c(R)$,
$\nu(R)$) for each $R$, were determined by the least squares
fitting. (Note that also $\nu(R)$ is a fitting parameter in our analysis.) In order
to estimate the statistical error, we performed the following
bootstrap procedure.  From several hundreds Robbins-Monro runs, the physical quantities
and the parameters [$\beta(L)$ and $\hs_\rmc(L)$] were obtained with
some statistical error bars. Then the data were resampled from the Gaussian
distribution with the estimated mean and variance.  The generated
samples were fitted for large-enough system-size data that
minimize $\chi^2/\mathrm{d.o.f}$, where the asymptotic
form~\eqref{extrapolate_A} would approximate the plots well (we used
the data for $L\in[24,64]$ and obtained $\chi^2/\mathrm{d.o.f.}\approx
2.0$).  This procedure was repeated 4000 times, which yielded as
many fitted functions. By taking the average of the function values,
we obtained a whole shape of $\hs_\rmc(L)$ that should be
asymptotically accurate, which is shown in Fig.\,\ref{h0_A_2.pdf} as
the solid line for each $R$.  Finally, the bootstrap estimation gave
the critical point $\hs_\rmc=0.99179(3)$ for $\hu=0$, where the number in the parenthesis denotes the standard error of the estimation in the last digit(s). It is consistent
with the previous report,
$\hs_\rmc=0.9919(4)$~\cite{PriyadarsheeCLB2006}, but more precise by
an order of magnitude.

It should be noted that in Fig.\,\ref{h0_A_2.pdf} the fastest convergence is achieved by the choice of $R=R_{XY}$, i.e.,
the leading correction seems $o(1/L^{1/\nu_{XY}})$ at $R=R_{XY}$. Meanwhile, the leading correction is likely in the order of $1/L^{1/\nu_{XY}}$ at $R=0.5$ and $0.7$, where
$\nu_{XY}=0.67155$~\cite{CampostriniHPRV2001} is the critical
exponent of the correlation length for the 3D-$XY$
universality class. These findings indicate that this phase transition belongs to the 3D-$XY$ universality class.

In Fig.\,\ref{h0_all.pdf}, we present the system-size dependence of the
inverse temperature, the static structure factor, and the transverse
susceptibility.
Assuming the asymptotic forms [Eqs.\,\eqref{dynamical_exponent}, \eqref{s_fac}, and \eqref{susceptibility}],
we conclude $z=0.992(6)$, $\gamma/\nu=1.967(6)$, and
$\theta=0.968(5)$. These results are fully consistent with the
scenario of the 3D-$XY$ universality class, $\gamma/\nu=1.9620(4)$ and $\theta=0.9620(4)$~\cite{CampostriniHPRV2001}, with $z=1$.
The final estimates for $z$, $\theta$, and
$\gamma/\nu$, which were evaluated from the asymptotic behavior of the different quantities, indeed satisfy the scaling relation
\begin{equation}
  \gamma/\nu = \theta + z
\end{equation}
within the error bar.

\begin{figure}[tbp]
\centering
\vspace{1em}
\includegraphics[width=80mm]{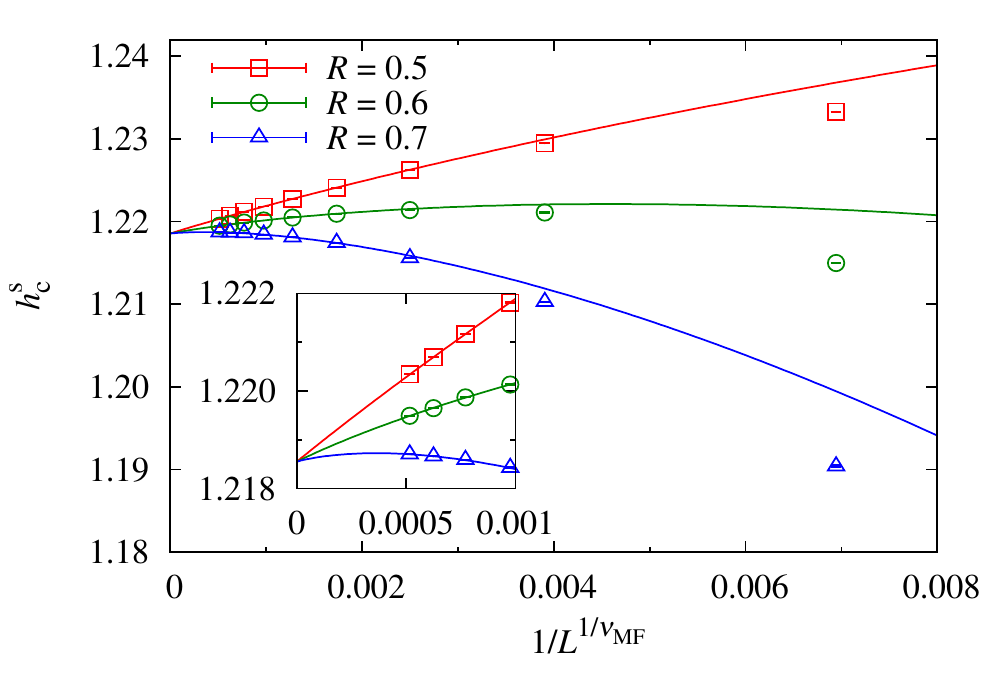}
\vspace{-1em}
\caption
  {(Color online) System-size dependence of the critical staggered field $\hs_\rmc(L)$ for $\hu=0.5$. The extrapolated value is $\hs_\rmc=1.21855(2)$. The lines were obtained by the same bootstrap approach with $\hu=0$ (see the body). The inset shows the detail of the extrapolation for larger $L$.}
\label{h0.5_A.pdf}
\end{figure}

\begin{figure}[tbp]
\centering
\vspace{1em}
\includegraphics[width=80mm]{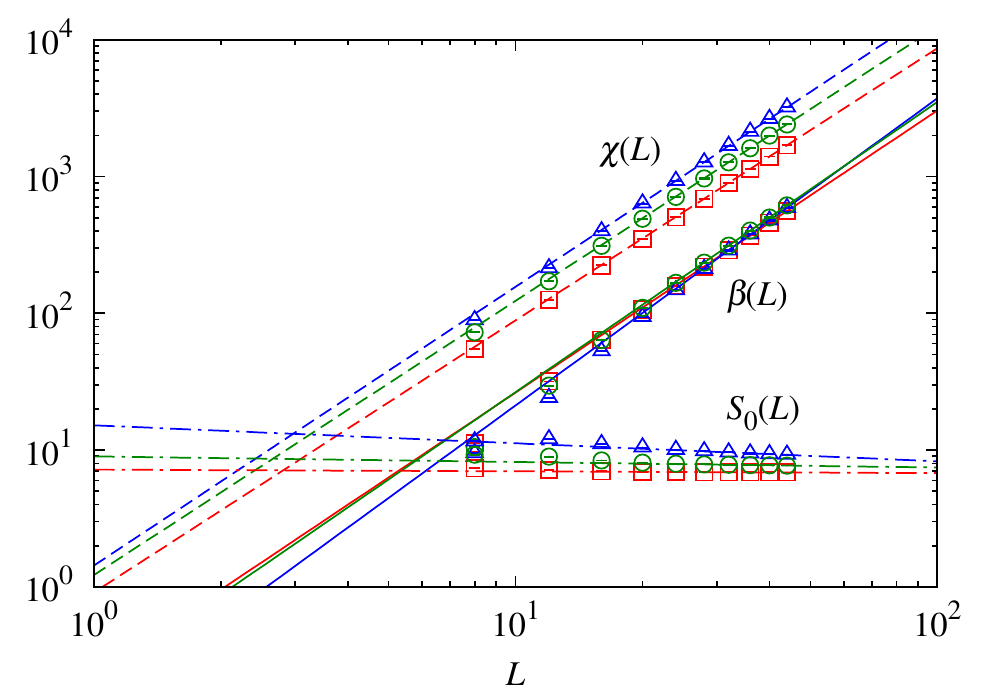}
\vspace{-1em}
\caption
    {(Color online) System-size dependence of the transverse susceptibility\,\eqref{def_sus}, the structure factor\,\eqref{def_sfac}, and the inverse temperature at $\hs_\rmc(L)$ for $\hu=0.5$. The red squares, green circles, and blue triangles are
  the data plots for $R=0.5$, $0.6$, and $0.7$, respectively.}
\label{h5_all.pdf}
\end{figure}

\begin{figure}
\centering
\includegraphics[width=80mm]{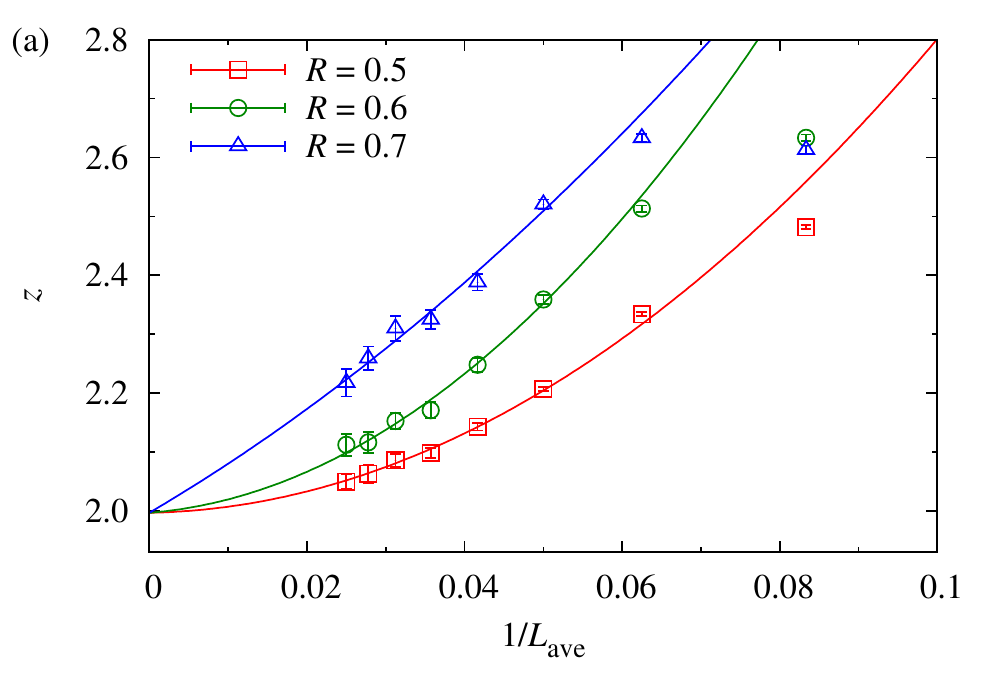}
\includegraphics[width=80mm]{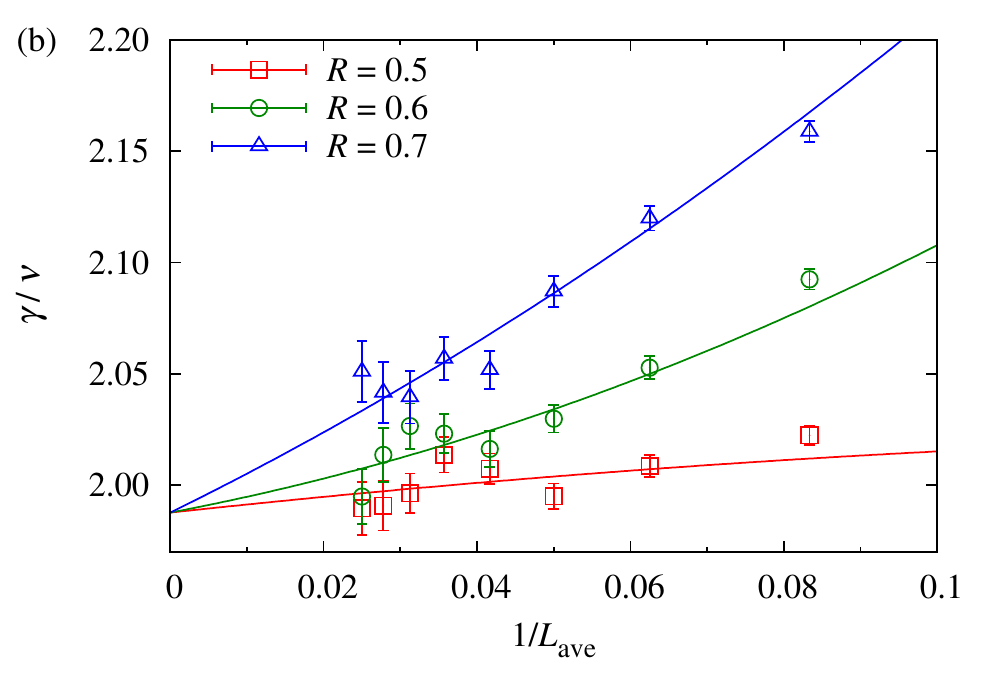}
\includegraphics[width=80mm]{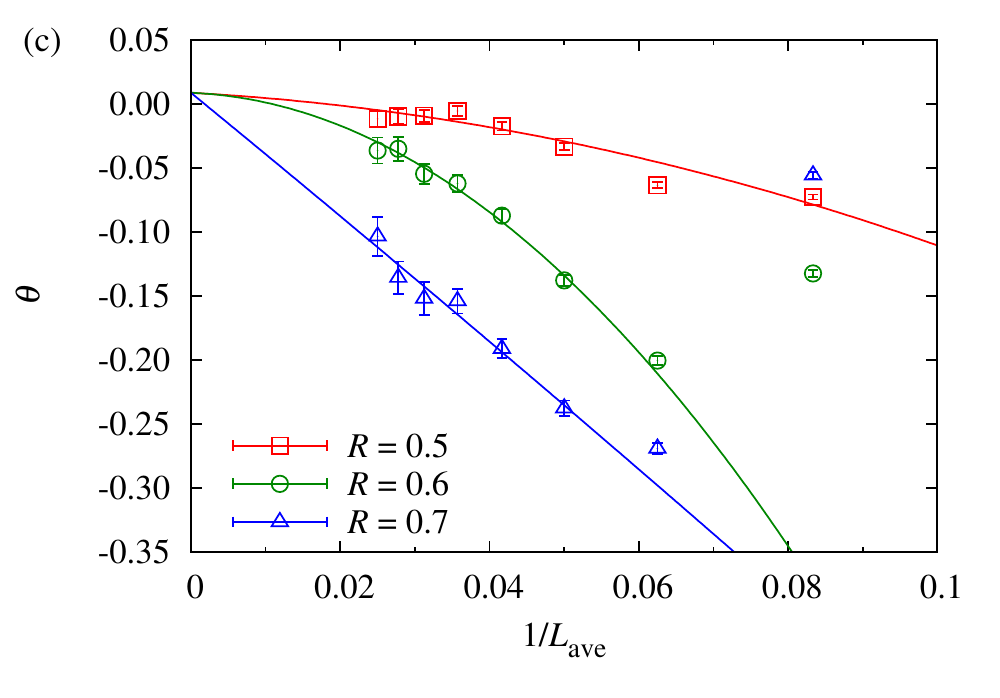}
\caption
  {(Color online) Finite-size corrections of the critical exponents: (a) the dynamical exponent, $z$, (b) the exponent of the susceptibility, $\gamma/\nu$, and (c) that of the structure factor, $\theta$. Each data point represents the exponent obtained by the fit for each triad (see the body).}
\label{exponents.pdf}
\end{figure}

\subsection{For $\hu=0.5$}

Next, we discuss the critical point and the critical exponents for the case with finite uniform magnetic field $\hu=0.5$.  In this case
we used $R=0.5$, $0.6$, and $0.7$ for $L=8,12,16,20,24,28,32,36,40,44$.  Following the same procedure with the case for $\hu=0$, we obtained $\hs_\rmc=1.21855(2)$, the quantum critical point. In the fitting procedure, the data with $L\in[24,44]$ were used ($\chi^2/\mathrm{d.o.f.}\approx 0.8$).

The system-size dependence of the physical quantities is shown in
Fig.\,\ref{h5_all.pdf}.  In comparison to the case with $\hu=0$
shown in Fig.\,\ref{h0_all.pdf}, larger corrections to
scaling are seen, especially for $\beta(L)$.  To cope with the strong
finite-size corrections, we took the following procedure: Assume we
have data points at system sizes $L=L_1$, $L_2$, $\cdots$, $L_n$,
where $L_1<L_2<\cdots<L_n$.  First, we construct triads consisting of
the data with three consecutive system sizes as $(L_1,L_2,L_3)$,
$(L_2, L_3, L_4)$, $\cdots$, and $(L_{n-2}, L_{n-1}, L_n)$, defining
$L_{\mathrm{ave}}$ as the average system size of each triad. Next, we
fit each triad with a simple power function, i.e.,
$y(L)=a(L_\mathrm{ave})\times L^{b(L_\mathrm{ave})}$ with
$a(L_\mathrm{ave})$ and $b(L_\mathrm{ave})$ the fitting parameters
depending on $L_\mathrm{ave}$. The error of $b(L_\mathrm{ave})$ is
estimated by the bootstrap method as explained above.  Then, we fit
$b(L_\mathrm{ave})$ with a quadratic function of $1/L_\mathrm{ave}$,
and extrapolate the critical exponent in the thermodynamic limit.  In
our fitting procedure of the exponent, we assume that the fitting function should be
monotonic.

The extrapolation results are shown in Fig.\,\ref{exponents.pdf}.
For the dynamical exponent, we estimated $z=2.00(2)$;
the effective dimension of the imaginary-time axis changes from one to two by the introduction of uniform magnetic field $\hu$. 
As for the other critical exponents,
$\gamma/\nu=1.99(1)$ and $\theta=0.01(1)$ were
obtained. These values coincide with the mean-field exponents, i.e.,
$\gamma/\nu=2$ and $\theta=0$. This is consistent with the result for the dynamical exponent, $z=2$, by which the effective dimension of the critical theory becomes four, the upper critical dimension. Thus we have demonstrated that our finite-size scaling method enables us to extract the dynamical exponent successfully without any prior knowledge of the value of $z$.  The universality of the quantum phase transition without particle-hole symmetry belongs to the mean-field universality class. This is consistent with the discussion
on the Bose-Hubbard model in Ref.\,\citenum{FisherWGF1989}.



\section{Summary and discussions}\label{summary}
In the present paper, we have presented the unified finite-size scaling method that
works well regardless of the value of the dynamical exponent. 
During the WLQMC simulation, the system size in the imaginary-time direction in the path-integral representation is adjusted automatically so as to satisfy the conditions,
$\xi/L=R$ and $\xi_\tau/\beta=R_\tau$, based on the Robbins-Monro stochastic approximation. This auto-tuning procedure guarantees that the coupling constant converges to the critical point and the inverse temperature is proportional to $L^z$ for large enough $L$.

We then applied the method to the two-dimensional $S=1/2$ quantum $XY$
model in uniform and staggered magnetic fields. The correlation
lengths were measured by the worm algorithm based on the continuous
imaginary-time representation.  In the absence of the uniform magnetic field,
$\hu=0$, our numerical results are consistent with the
3D-$XY$ universality class.  This system can be mapped into the
half-filled hard-core boson system.  The Lorentz invariance, $z=1$, reflects the particle-hole
symmetry at half-filling.  In the case with $\hu=0.5$, we have concluded
that the dynamical exponent changes to two and the other exponents
take the mean-field values. This result of the mean-field
universality is consistently explained by the conclusion that the
dimension of the effective field theory is four;
$d+z=4$, the upper critical dimension.  Our conclusion $z=2$ agrees on the discussion in
Ref.\,\citenum{FisherWGF1989}, in which the authors claimed that the
fourth-order term in the effective action becomes irrelevant.

The method proposed in the present paper is applicable also to models with
randomness. For example, the method will be effective for systems 
whose dynamical exponent depends on the parameters, such as the random Ising 
model in random transverse field~\cite{PichYRK1998,RiegerK1999}, where the dynamical exponent may take an irrational value or even becomes infinite~\cite{Fisher1995}. It is thus extremely difficult to analyze the properties of quantum criticality by the conventional strategies.
By using our method, one does not need any assumption about the dynamical exponent, which should be quite effective for the systematic investigation of the randomness-driven quantum critical point causing extreme space-time anisotropy.

\section*{Acknowledgments}
The simulation code used in the present study has been developed based on an open-source implementation of the worm algorithm~\cite{WORMSweb} using the ALPS Library~\cite{ALPS2011s, ALPSweb} and the BCL (Balance Condition Library)~\cite{SuwaT2010, BCLweb}.  The authors acknowledge the support by KAKENHI (No.\,23540438, 26400384) from JSPS, the Grand Challenge to Next-Generation Integrated Nanoscience, Development and Application of Advanced High-Performance Supercomputer Project from MEXT, Japan, the HPCI Strategic Programs for Innovative Research (SPIRE) from MEXT, Japan, and the Computational Materials Science Initiative (CMSI). S.Y. acknowledges the financial support from Advanced Leading Graduate Course for Photon Science (ALPS).

\appendix
\section{Convergence by the Robbins-Monro algorithm}\label{RM_convergence}
\begin{figure}
\centering
\includegraphics[width=80mm]{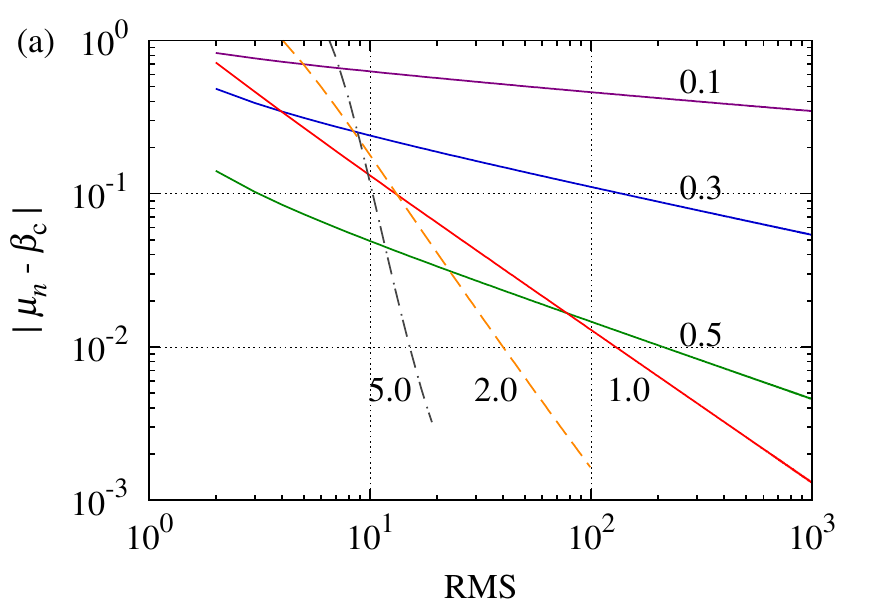}
\includegraphics[width=80mm]{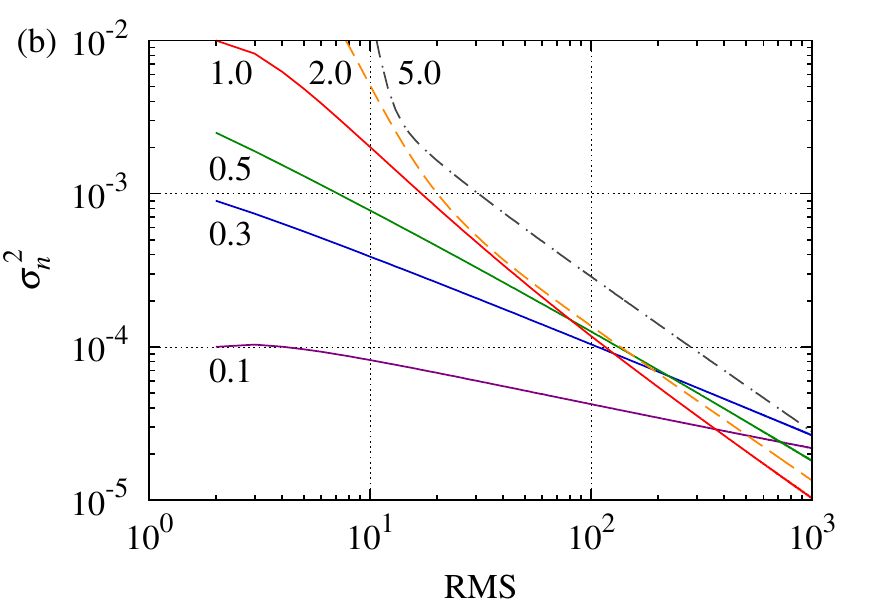}
\caption
  {(Color online) Convergence of the (a)~bias and (b)~variance for solving
    $f(\beta)=\exp(\beta)-1=0$ with the initial condition
    $\beta^{(1)}=\beta_\mathrm{init}=1$. At the $n$-th RMS, a value
    (corresponding to the relative correlation length in the
    quantum-phase-transition analyses) is generated from the normal
    distribution, $\mathcal{N}(f(\beta^{(n)}), 0.1)$.  Note
    that $a=1$ in this case. The results for
    $p=0.1$, $0.3$, $0.5$, $1.0$, $2.0$, and $5.0$ are shown
    (calculated from $10^8$ Robbins-Monro processes).  As shown in the
    text, the variance is minimized by the choice of $p=1/a=1$ at large enough RMSs, while the bias convergence is accelerated monotonically with respect to $p$.
  }
\label{mu_sigma.pdf}
\end{figure}

In this appendix, the time evolution of the probability distribution function driven by the Robbins-Monro iteration [Eq.\,\eqref{update}] is discussed.  Let us consider the situation in which the
physical quantity $A(\beta)$ is obtained by Monte Carlo simulation and the
distribution function of $A(\beta)$ is given by a Gaussian (normal) distribution written as
\begin{align}
P(A)&=\mathcal{N}(f(\beta),\sigma^2)\nonumber\\
&=\frac{1}{\sqrt{2\pi\sigma^2}}\exp\left[-\frac{(A-f(\beta))^2}{2\sigma^2}\right].
\end{align}
Also, we assume that $f(\beta)$ can be expanded as
\begin{equation}
f(\beta)\approx a(\beta-\beta_\rmc),
\end{equation}
near $\beta_\rmc$, the zero of $f(\beta)$.
Then, the asymptotic recursion relation between the distribution functions of $\beta^{(n)}$ and $\beta^{(n+1)}$ is written as
\begin{align}
&P_{n+1}(\beta^{(n+1)})\propto\int\diff\beta^{(n)}P_n(\beta^{(n)})\nonumber\\
&\times\exp\left\{-\frac{1}{2\sigma^2}\left[\frac{n}{p}(\beta^{(n)}-\beta^{(n+1)})-a(\beta^{(n)}-\beta_\rmc)\right]^2\right\}.\label{P_n+1}
\end{align}
In our procedure, we start from an initial condition $P_1(\beta^{(1)})=\delta(\beta^{(1)}-\beta_\mathrm{init})$.
Here we assume that the distribution function of $P_n(\beta^{(n)})$
will be approximated by a Gaussian for large enough $n$. Then the recursion relations for
the mean, $\mu_n$, and the variance, $\sigma_n^2$, of
$P_n(\beta^{(n)})$ are obtained as
\begin{align}
&\mu_{n+1}=\left(1-\frac{ap}{n}\right)\mu_n+\frac{ap}{n}\beta_\rmc\label{mu_n+1}\\
&\sigma_{n+1}^2=\sigma^2\left(\frac{p}{n}\right)^2+\sigma_n^2\left(1-\frac{ap}{n}\right)^2\label{sigma_n+1},
\end{align}
respectively. Eq.\,\eqref{mu_n+1} can be rewritten as
\begin{equation}
  \mu_{n+1} - \beta_\rmc = \left( 1 - \frac{ap}{n} \right) \left( \mu_n - \beta_\rmc \right) \label{mu_n}.
\end{equation}
The absolute value of $( 1 - ap / n )$ in Eq.\,\eqref{mu_n} is less than 1 for
sufficiently large $n$. Then $\mu_n \rightarrow \beta_\rmc$ for $n\to \infty$. Similarly, it can be proved that $\sigma_n^2\to 0$ for $n\to\infty$.

Next, let us
assume the leading term of $\sigma_n^2$ as ${s^2}/{n^{\alpha_s}}$, where $\alpha_s$ is an unknown constant and
$s^2$ is the asymptotic variance of $\sigma_n^2$.  From the approximation $(n+1)^{-\alpha_s}\approx n^{-\alpha_s}(1-\alpha_s/n)$ for $n\gg 1$, 
the lowest-order terms in Eq.\,\eqref{sigma_n+1} are evaluated as
\begin{equation}
-\frac{s^2\alpha_s}{n^{\alpha_s+1}} \approx \frac{\sigma^2p^2}{n^2}-\frac{2s^2ap}{n^{\alpha_s+1}}.
\end{equation}
When $\alpha_s < 1$, $1/n^{\alpha_s+1}$ dominates over $1/n^2$. Then $\alpha_s = 2ap$ and
\begin{equation}
  \sigma_n^2\sim\frac{1}{n^{2ap}}  \qquad \text{for $ap < \frac{1}{2}$.}
  \label{sigma_n2}
\end{equation}
When $\alpha_s = 1$, on the other hand,
\begin{equation}
\sigma_n^2\approx\frac{s^2}{n}=\frac{1}{n}\frac{\sigma^2 p^2}{2ap-1} \qquad \text{for $ap>\frac{1}{2}$.}\label{sigma_n}
\end{equation}
There is no solution under the assumption as for the case where $ap=1/2$, but it is expected that only some correction from $\sigma_n^2 \sim 1/n$ will appear.

Similar discussion holds for the mean of distribution, $\mu_n$. Assuming $\mu_n -
\beta_\rmc = k/n^{\alpha_m}$, with some constants $k$ and $\alpha_m$, we obtain
\begin{equation}
\beta_\rmc+\frac{k}{n^{\alpha_m}}\left(1-\frac{\alpha_m}{n}\right)=\left(1-\frac{ap}{n}\right)\left(\beta_\rmc+\frac{k}{n^{\alpha_m}}\right)+\frac{ap}{n}\beta_\rmc
\end{equation}
from Eq.\,\eqref{mu_n+1}. This results in $\alpha_m=ap$, and thus we have
\begin{equation}
  \mu_n-\beta_\rmc\sim\frac{1}{n^{ap}}   \qquad \text{for $ap > 0$.}
  \label{mu_n2}
\end{equation}

According to the asymptotic forms \eqref{sigma_n2}, \eqref{sigma_n}, and
\eqref{mu_n2}, let us discuss the dependence of the final error on the
number of Monte Carlo steps. It is obvious from Eqs.\,\eqref{sigma_n2} and \eqref{sigma_n} that it is better to set the gain $|p|$ large
enough so that $ap > 1/2$ is satisfied. Otherwise the convergence of
the variance becomes slower. We
will call one iteration of the Robbins-Monro feedback process
[Eq.\,\eqref{update}] a Robbins-Monro step (RMS), and suppose that a
whole calculation consists of $N_\mathrm{R}$\,RMSs and each RMS
has $N_{\mathrm{M}}$ Monte Carlo updates. The total computational
cost is proportional to $N_{\mathrm{tot}}\equiv N_\mathrm{R}\times
N_\mathrm{M}$.  From Eq.\,\eqref{sigma_n}, the variance of the
estimate after $N_\mathrm{R}$ RMSs is given by
$\sigma_{N_\mathrm{R}}\approx
s^2/N_{\mathrm{R}}\sim\sigma^2/N_\mathrm{R}$ for $ap > 1/2$.  Using
$\sigma^2\approx s^2_{\mathrm{MC}}/N_\mathrm{M}$, where
$s^2_\mathrm{MC}$ is the asymptotic variance of the Monte Carlo
estimation, we obtain
\begin{equation}
\sigma_{N_\mathrm{R}}\sim \frac{s^2_\mathrm{MC}}{N_\mathrm{tot}}.\label{cost}
\end{equation}
This means that the asymptotic variance depends only on the total number of Monte Carlo updates $N_{\mathrm{tot}}$.

Next, let us discuss the optimal choice of the gain $p$. For $ap>1/2$, the convergence of $\mu_n$ is faster than that of $\sigma_n$. Thus, we should minimize $\sigma_n$ in Eq.\,\eqref{sigma_n}; then we derive
\begin{equation}
  p=\frac{1}{a}.
\end{equation}
Fig.\,\ref{mu_sigma.pdf} shows the $p$ dependence of $\mu_n$ and $\sigma_n^2$ calculated from an exemplary case:
\begin{equation}
  f(\beta) = \exp(\beta)-1,
\end{equation}
where $\beta_\rmc = 0$ and $a=1$.
While the convergence of the mean becomes faster monotonically as $p$ increases, the variance takes a minimum value at $p=1/a=1$ and increases again for larger $p$.
In the simulation presented in the main text, we performed short preparatory calculations for small systems in order to roughly estimate $a\approx a^*$, and set $p=1/a^*$ for succeeding long runs. We performed several hundreds of Robbins-Monro iterations, each of which has $N_\mathrm{M}=500$ worm updates.


\section{Measurement of off-diagonal correlation in the worm algorithm}\label{worm}
We explain the way to measure off-diagonal correlation in the worm algorithm.
Let us begin with measuring the static structure factor defined in Eq.\,\eqref{def_sfac}. This quantity is
easily rewritten by the spin raising and lowering operators. We then need to evaluate the thermal average,
\begin{equation}
\Big\langle \sum_{j,k}S_j^+S_k^- \Big\rangle = \frac{\mathrm{Tr}\Big[\sum_{j,k} S_j^+S_k^-\mathrm{e}^{-\beta\mathcal{H}}\Big]}{\mathrm{Tr}\,\mathrm{e}^{-\beta\mathcal{H}}}. \label{Spm}
\end{equation}
The numerator and denominator in
Eq.\,\eqref{Spm} correspond to the partition function of extended world-line configurations with a worm and those of the original world-line configurations, respectively. In other words, Eq.\,\eqref{Spm}
can be read as the frequency of events that the worm head visits the same
imaginary time with the tail.  We thus can measure the static structure factor by simply
counting the frequency of extended configurations that contribute to the numerator in Eq~.\eqref{Spm} during each worm update.

We have also evaluated the dynamic structure factor at imaginary frequency $i\omega$.  It is given by the canonical
correlation function as
\begin{align}
C(\vec{q},i\omega) &= \frac{1}{L^2\beta} \Big\langle
\sum_{j,k}\int_0^\beta\!\! \mathrm{d}\tau_1\int_0^\beta\!\!\mathrm{d}\tau_2\,S_j^+(\tau_1)S_k^-(\tau_2)\nonumber\\
&
\times \exp\left\{-i\left[\omega(\tau_2-\tau_1)+\vec{q}\cdot(\vec{r}_k-\vec{r}_j)\right]\right\}\Big\rangle,\label{dst}
\end{align}
where $i\omega=2\pi i/\beta$ is the lowest Matsubara frequency in our simulation.
The spatial phase factor $\mathrm{e}^{-i\vec{q} \cdot (\vec{r}_k -
  \vec{r}_j)}$ can be calculated and stored in advance. Then, when the
head moves in the worm-update process, the imaginary-time integral is
evaluated. In simulations, every time the head reaches a bond operator, a part of the imaginary-time
integral is performed.  Since the $\tau$-dependent part of the
integrand is simply given by $\mathrm{e}^{i\omega\tau}$, we can evaluate the part of the
integral exactly at each head move. The spatial phase factor is multiplied (if necessary). The matrix
elements of the head and tail also need to be considered (they are simply
one in the case with $S=1/2$). The contribution to the integral at each head
move is summed up until the head returns back to its tail. The average value of the
summed integral for each worm insertion will provide the target
quantity~\eqref{dst}.

The transverse susceptibility~\eqref{def_sus} is expressed as $\chi=C(\vec{0},0)$. The
correlation length can be estimated by the second-moment
method~\cite{CooperFP1982,TodoK2001}; the correlation length in
the $x$ direction, $\xi_x$, is expressed as
\begin{equation}
\xi_x = \frac{1}{|\delta\vec{q}_x|}
\sqrt{\frac{C(\vec{q}_0,0)}{C(\vec{q}_0 + \delta\vec{q}_x, 0)}-1},
\end{equation}
where $\delta\vec{q}_x=(2\pi/L,0)$ and $\vec{q}_0=(0,0)$. Similarly, the
correlation length in the imaginary-time direction, $\xi_\tau$, as
\begin{equation}
\xi_\tau = \frac{1}{\omega}
\sqrt{\frac{C(\vec{q}_0,0)}{C(\vec{q}_0, i\omega)}-1}.
\end{equation}

\bibliography{main}
\end{document}